\documentclass[11pt]{article}
\usepackage{graphicx}

\newcommand{\fig}[1]{Fig.~\ref{fig.#1}}

\newcommand{\tbl}[1]{Table~\ref{table.#1}}

\newcommand{\figlabel}[1]{\label{fig.#1}}
\newcommand{\tbllabel}[1]{\label{table.#1}}

\newcommand{\figwidth}{3in}

\newcommand{\nBidders}{\ensuremath{n}} 

\begin{document}

\title{Experiments with Probabilistic Quantum Auctions}
\author{Kay-Yut Chen \and Tad Hogg}
\maketitle

\begin{abstract}
We describe human-subject laboratory experiments on probabilistic
auctions based on previously proposed auction protocols involving
the simulated manipulation and communication of quantum states.
These auctions are probabilistic in determining which bidder wins,
or having no winner, rather than always having the highest bidder
win. Comparing two quantum protocols in the context of first-price
sealed bid auctions, we find the one predicted to be superior by
game theory also performs better experimentally. We also compare
with a conventional first price auction, which gives higher
performance. Thus to provide benefits, the quantum protocol requires
more complex economic scenarios such as maintaining privacy of bids
over a series of related auctions or involving allocative
externalities.
\end{abstract}

\section{Introduction}

Auctions usually have a well-defined decision rule to determine
which bidder, if any, wins the auction. In these cases, the outcome
is a deterministic function of the submitted bids. More generally,
an auction may include various informal or loosely defined
preferences, such as a diversity of winners to avoid too much
concentration in a single supplier, an extreme bid suggesting a low
quality supplier, or public policy requirements~\cite{klemperer06}.
To the extent these factors are not clearly expressed to
participants or are not fully evaluated until after the auction,
bidders face some uncertainty on how the auction winner will be
selected, e.g., whether the highest bidder will win.

While such additional criteria can be complicated, an approach to
understanding their effect on bidder behavior is to treat the
auction outcome as probabilistic. That is, while traditional auction
criteria, such as the amount bid, affect the winner selection, the
relation is probabilistic rather than deterministic.

More speculative examples of probabilistic auctions are quantum
auctions~\cite{harsha06,guha07}, where bidders use quantum states to
express their bids. Such auctions use quantum information
processing~\cite{chuang00} both to express the bids and determine
the winner(s). A potential benefit of quantum auctions is helping to
preserve privacy of the losing bids: the observation at the end of
the protocol to determine the winner also destroys the quantum state
encoding the bidders' behavior. Cryptographic methods can also hide
information~\cite{naor99} through the presumed computational
difficulty of breaking the code. However, cryptographic approaches
retain the information after completing the protocol, which may be
revealed if the corresponding keys become available, either
accidentally or intentionally. Moreover, the political context of
some auctions can lead to difficulties in using economically
efficient methods when information, beyond just the winner and
price, remains available~\cite{klemperer06}. Quantum auctions could
provide an attractive alternative to address these concerns. Quantum
auctions, through using entangled states, also offer a
privacy-preserving approach to scenarios in which one bidder cares
about what another bidder wins (called an ``allocative
externality''), and allow compact expression of bids for
combinatorial auctions~\cite{cramton06}.

Other examples of quantum economic mechanisms include encouraging
cooperation in the context of the prisoner's
dilemma~\cite{eisert98,eisert00,du01,du02},
coordination~\cite{huberman03,mura03}, the minority
game~\cite{flitney07} and public goods provisioning~\cite{chen03}.
In particular, a quantum public-goods mechanism can significantly
reduce the free-rider problem without a third-party enforcer or
repeated interactions, both in theory and practice~\cite{chen06}.
Technology for manipulating and communicating just a few qubits
could be sufficient to implement such mechanisms. The use of quantum
information for economic mechanisms contrasts with the attention
given to computational advantages~\cite{shor94,grover96} of quantum
computers with many qubits, which are much more difficult to
implement physically than few-qubit economic applications.

A key question for economic mechanisms is how well they perform. In
this paper we consider this question for a quantum auction
protocol~\cite{harsha06} that mimics the conventional first-price
sealed-bid auction. In this type of conventional auction,
participants each submit a single bid, without knowledge of the bids
submitted by others. This contrasts with the more common open outcry
auctions where bidders learn each others bids and can choose to
increase their bid if someone else bids higher. Moreover, in
first-price auctions, the winner is the bidder submitting the
highest bid, and the winner pays the amount of that bid. We focus on
the first-price sealed-bid auction due to its simplicity.

A core issue is whether the probabilistic auction protocol, with the
appropriate design, performs as well as an efficient deterministic
auction of the same items. A partial game theoretic analysis of this
quantum auction indicates how ideal rational players would react to
the probabilistic nature of the outcomes~\cite{harsha06}. In
particular, this analysis indicates design details of the
distributed quantum search to find the winner affect the game theory
incentives and the resulting auction efficiency.

However, there are many reasons to be skeptical of game theory
predictions of human decision-making behavior. One particularly
relevant issue is that standard theory has difficulty predicting how
people make decisions under uncertainty~\cite{kahneman79,camerer95}.
In the quantum auction, bidders face three different types of
uncertainty. First, they do not know other bidders' valuation of the
item. Second, they do not know how other bidders will bid as a
function of their values, a form of ``strategic uncertainty'' that,
in some cases, is useful in encouraging more participation in the
auction than would be the case for rational
bidders~\cite{klemperer06}. To some extent, game theory assumes away
this strategic uncertainty by having everyone use the equilibrium
strategy. The last type of uncertainty, unique to probabilistic
auctions, comes from the ability of the bidders to change winning
probabilities by manipulating the auction protocol. That is, the
winning probabilities are not exogenously given values, but instead
determined by the bidders' behaviors. Other situations often leading
to poor predictive performance of game theory include situations
with multiple equilibria or mixed strategy equilibria, where
rational players select among their available choices according to a
probability distribution. Moreover, in some cases people may use
non-equilibrium strategies that are more Pareto efficient than the
Nash equilibrium, e.g., cooperating to some extent in single
prisoner's dilemma situations. These features often occur in quantum
games, including the quantum auction protocol evaluated in this
paper. Thus, it is important to test quantum auctions with real
human bidders, rather than relying only on game theory analyses, to
evaluate their suitability for real world applications.

In this paper, we report a series of economic experiments, using
human subjects, designed to test whether people can perform
successfully in quantum auctions. Our experiments compare the two
existing protocols for quantum auctions~\cite{harsha06}.
Furthermore, we test whether game theory predicts how people bid in
quantum auctions. Since we examine human behavior in a controlled
laboratory setting, it is sufficient to simulate the behavior of the
quantum auctions with conventional computers for our experiments.
Thus we can evaluate the performance and potential usefulness of
economic applications relying on few-qubit quantum information
processing prior to their physical implementation.

The remainder of this paper describes a quantum auction protocol,
our experiment design and the results comparing both conventional
deterministic auctions and probabilistic auctions based on the
quantum protocols simulated with conventional computers.

\section{A Quantum Auction Protocol}

The quantum auction~\cite{harsha06} uses quantum superpositions to
represent bids and adiabatic quantum search~\cite{farhi01} to
identify the winning bid. The overall procedure for the bidders and
auctioneer is as follows. The auctioneer starts by creating a set of
quantum physical systems in a specified initial state. This set has
one system for each bidder, which consists of a number of qubits.
The protocol proceeds through a series of rounds which implement a
distributed version of an adiabatic search. In each round, the
auctioneer starts with all the quantum systems, performs an
operation on them and then sends one system to each bidder. Each
bidder performs an operation on the system received from the
auctioneer and then returns it to the auctioneer. After completing
all the rounds, the auctioneer measures all the quantum systems to
obtain a unique outcome for the auction while simultaneously
destroying the bid states. Thus the bids of the losing bidders are
never revealed. This protocol includes outcomes in which none of the
bidders win, which is analogous to conventional auctions with a
reservation price: if no one bids at least that much, the seller
keeps the item.

Each bidder privately selects their operator. Nominally, this bidder
operator corresponds to the bid they wish to place according to an
encoding announced prior to the auction. In this case, after
sufficiently many rounds, the adiabatic theorem gives high
probability that when the auctioneer measures the final state of the
quantum systems, the observations will indicate the highest bid and
the corresponding bidder. This outcome arises through the choice of
operators and initial states for the quantum systems. Specifically,
the bidders' operators place the quantum systems in the ground state
of an initial Hamiltonian. During the subsequent rounds, the
auctioneer gradually changes the operation performed until the final
Hamiltonian corresponds to the value of each outcome to the
auctioneer, with the ground state having the highest value.
Typically this value is the revenue received by the auctioneer.

Bidder operators constrain the state to only have nonzero amplitudes
among outcomes bidders are willing to pay, so the adiabatic search
finds the highest-revenue state consistent with the bidders'
choices. This process can be viewed as an adiabatic search
constrained to a subspace of all possible outcomes rather than
finding the global state of maximum revenue for the auctioneer,
which would amount to asking bidders to pay the maximum possible
price.

This protocol has two distinct reasons for probabilistic outcomes.
First, the auctioneer may not perform enough rounds to satisfy the
adiabatic theorem, in which case amplitude may spread among various
eigenstates and lead the final observation to give outcomes other
than the ground state. Second, a bidder may strategically choose an
operator different from that corresponding to the intended bid. Such
choices can form an initial state for the adiabatic search that is
not the ground state of the initial Hamiltonian, but instead some
superposition of eigenstates of the initial Hamiltonian. In this
case, a slow adiabatic search will produce a corresponding mixture
of eigenstates of the final Hamiltonian, resulting in probabilistic
outcomes for the auctioneer's measurement. In particular, such
choices could give some probability for low bidders to win.

Our experiments consider the choices bidders make, so we focus on
this second source of probabilistic outcomes. Thus we take the
auctioneer to use enough rounds to satisfy the adiabatic theorem
with probability close enough to one that outcomes violating the
adiabatic theorem do not occur in our limited set of experiments.

A single bidder, having only one of the quantum systems to operate
on, cannot form arbitrary initial states of the full set. Thus an
important consideration for the auction design is the range of
excited states of the Hamiltonian available to a single bidder. Game
theory suggests simple changes in the design of the Hamiltonian (as
implemented by the auctioneer) significantly affects the strategic
behavior of players by changing the available excited states to a
single bidder~\cite{harsha06}. Testing the effect of such changes on
actual human decision-makers is the key focus of the experiments
described below.

\section{Economic Experiments}

Game theory, along with traditional economic theory, is the standard
theory for human decision-making in economic contexts. This theory
makes strong assumptions of rationality, including: a) each
decision-maker is selfish and maximizes his or her preferences, b)
they make no mistakes, and c) they know that all the other
decision-makers are rational. While these assumptions are not
accurate descriptions of how people actually make decisions, game
theory provides key insights into strategic decision-making behavior
with fairly accurate predictions in some contexts.

Since Chamberlain~\cite{chamberlain48} and Smith~\cite{smith62},
there has been growing interest in using economic
experiments~\cite{kagel95} to understand the behavioral factors
affecting decisions. These factors include, but not limited to,
those arising from social preferences (how people deal with each
other), individual bounded rationality (how people make mistakes in
decisions) and uncertainty~\cite{kahneman79}. In an experiment,
human subjects are recruited to play the roles of decision makers
and receive monetary rewards based on their profits during the
experiment. Subjects receive a full description about the experiment
with no deception. Anonymity with respect to roles and payment is
preserved.

While no one expects subjects to engage in complicated mathematical
calculations during experiments, there is ample evidence that
subjects with adequate training and instructions can make good
decisions within a laboratory setting where they only have limited
amounts of time. For example, subjects respond strategically to
nonlinearities in a first price auction when given about one minute
to make their bidding decisions~\cite{cox88,chen98}. These auction
experiments involve bidding decisions that are a bit less
complicated than the experiments we report in this paper, which
allowed a bit more time for each decision. As another point of
comparison, an experiment with much more complicated decisions than
our experiments allowed subjects 5 to 10 minutes for each
decision~\cite{charness02}. Moreover, subjects without specific
domain knowledge can make similar strategic choices to those trained
to make these choices, as examined, for example, with the complex
task of allocating multiple resources on the space
station~\cite{plott96}.

Previous work has also addressed how subjects learn about relevant
aspects of the game through interacting with the software as opposed
to being given a precise mathematical description. One example using
decision support to convey key aspects of a complex game is
experiments with electricity markets~\cite{chen07b}. Another example
is an experiment with a manufacturing system with complicated demand
function~\cite{chen08}. The demand function was only shown to the
subjects through a software decision support tool. No explicit
formulae or tables were given. The authors found that participants
made effective decisions based only on these tools and thus
successfully identified behavior patterns driven by changes in
contract policies. This approach using ``what if'' scenario tools is
similar our method of instructing subjects about the probabilities
of winning the auctions in the experiments reported in this paper.

Economic experiments have limitations, primarily with regard to
involving small groups of people over short times (typically a dozen
people making decisions over a period of a few hours). Thus some
extrapolation is required from experimental results to behavior of
real economic institutions. Nevertheless, such experiments provide a
useful addition to game theory and empirical observations of
economic institutions. In particular, laboratory experiments involve
both real human decision-making and control over economic variables
(such as supply and demand, and the information exchanged among
participants). Game theory lacks the former property and empirical
observations of existing institutions lack the latter. Because of
the limitations of experiments and game theory, we cannot expect
them to always give quantitative predictions for how large groups of
people would behave in real-world situations. Instead experiments,
by incorporating behavioral effects of real decision making, can
indicate how different mechanisms likely will compare, e.g., which
will likely work better. Such comparisons are sufficient to help
decide among competing mechanism choices, e.g., when designing
auctions.

\section{Experiment Design}

We conducted economics experiments to study actual behavior with
probabilistic auctions. We used auctions with exactly three bidders
and simulated the quantum auction protocol with conventional
computers so the probabilities of each outcome in the experiment
corresponded to that of the auctions with the given bidder inputs.

Each experiment consisted of a number of periods. At the beginning
of each period, subjects were grouped randomly into groups of three
and each group bid in their own auction. We determined the value for
the auctioned item for each bidder by randomly generating a value,
between 0 and 100, for each participant. Bidders then entered their
choices and the auction outcome was revealed. If a person won the
auctioned item, he or she would receive as profit the difference
between their value for the item and the amount of their bid. The
groupings and bidder values were regenerated for each subsequent
period. At the end of the experiment, each person was paid the total
profit they had received in all periods (converted to dollars at a
preannounced exchange rate).

\subsection{Experimental Treatments}

Our experiments compared three treatments:
\begin{enumerate}

\item Classical First Price Sealed Bid Auction

This classical auction has been well studied both theoretically and
experimentally. We included this treatment as a benchmark because it
is efficient in both theory and practice. In this treatment, in each
period, each subject entered a bid into the computer. The highest
bid, in each group of three, won the auction.

\item Quantum Auction with standard search\label{standard}

In this quantum auction, each subject entered a bid, as well as two
numbers we dubbed ``x'' and ``y'' for the auction. $x$ and $y$ are
two real numbers in the range from 0 to 12, with 12 equivalent to 0,
as on the face of a clock. Unknown to the subjects, these two
numbers determine a quantum operator specifying their bids for the
initial state of a subsequent distributed quantum search to find the
winner~\cite{harsha06}, using the standard quantum adiabatic search
procedure~\cite{farhi01}.

The subjects saw an auction that requires each bidder to specify 3
numbers (a bid, $x$ and $y$) instead of just one number (the bid) as
in the classical auction. Furthermore, as opposed to the highest bid
always winning, lower bids also have the possibility of winning
depending on the values of $x$, $y$ and the bid of every bidder.
Thus, this is a probabilistic auction with the subjects having some
control over the probabilities. Moreover, there is also a chance
that no one wins.

\item Quantum Auctions with permuted search

This auction is exactly the same as treatment~\ref{standard} except
that the probability of winning has a different dependency on the
bids, $x$ and $y$ values, corresponding to a permuted search with
better performance according to game theory~\cite{harsha06}. If
every bidder in an auction sets $x$ to 0 and $y$ to 3, then the
probability of the highest bid winning is 1, and this is a Nash
equilibrium~\cite{harsha06} at least with respect to changes in the
initial state.

\end{enumerate}

\fig{prob lowest wins} illustrates a difference between the standard
and permuted search methods by showing the probability the
\emph{lowest} bid wins as a function of the $x$ and $y$ values that
one bidder selects when the other two bidders select $x=0$ and
$y=3$. When all three bidders select $x=0$ and $y=3$, the highest
bidder always wins. With the standard search, in this situation each
bidder is tempted to change values to one of the peaks in this
figure (e.g., $x=6$ and $y=3$) giving a high chance to win with a
low bid, and hence earn a large profit from the auction. However, if
the other bidders also make this choice then there is always no
winner for the auction. By contrast, the permuted search method does
not have this temptation: when the other bidders select $x=0$ and
$y=3$, there is no probability for either the lowest bid or the
second lowest bit to win no matter what the remaining bidder
chooses.

\begin{figure}
\begin{center}
\includegraphics[width=\figwidth]{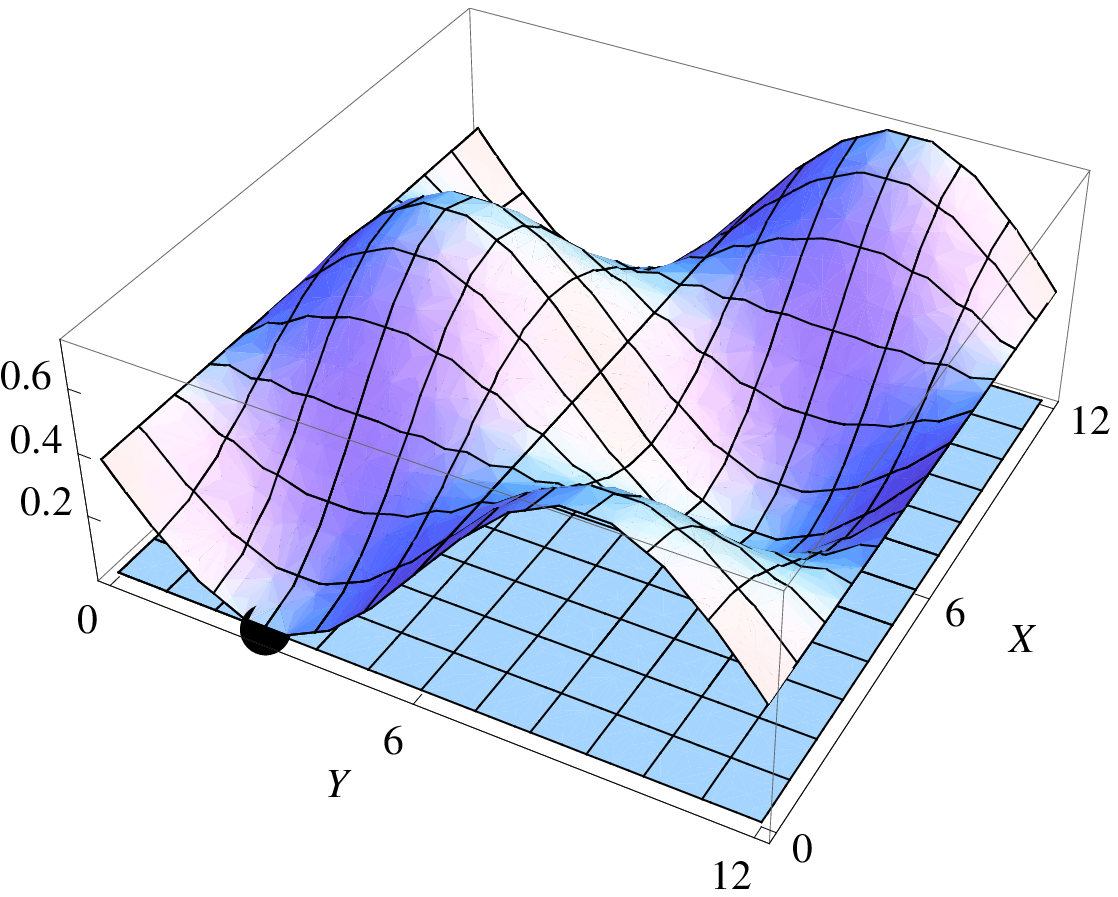}
\end{center}
\caption{\figlabel{prob lowest wins}The probability the lowest of
three bidders wins as a function of $x$ and $y$ values selected by
that bidder when the other two bidders select $x=0$ and $y=3$
(values indicated by the large black point). The upper surface is
for the standard search and the lower surface (with zero probability
for the lowest bidder to win) is for the permuted search.}
\end{figure}

Each experiment session was divided into two or three treatments as
summarized in~\tbl{summary}. That is, we ran a series of periods
using one treatment and then switched to a different treatment. The
subjects were notified before each new treatment took effect. We
report a total of three experiments. Two were conducted with
Stanford students covering all three treatments. The remaining
experiment (number 2) was conducted with physicists from HP Labs and
only two of three treatments were used due to time constraints.

\begin{table}[t]
\begin{center}
\begin{tabular}{cclc}
experiment & number of subjects & treatment & number of periods \\
\hline
1  & 12             & classical         & 35\\
                &   & quantum standard  & 35\\
                &   & quantum permuted  & 31\\ \hline
2  & 6              & classical         & 30\\
                &   & quantum permuted  & 40\\ \hline
3  & 12             & classical         & 30\\
                &   & quantum standard  & 30\\
                &   & quantum permuted  & 30\\ \hline
\end{tabular}
\end{center}
\caption{Summary of the experiments. The multiple treatments in each
experiment were done in the order listed.}\tbllabel{summary}
\end{table}

\subsection{Subjects' View of the Experiments}

We provided experiment instructions to the subjects via the
web\footnote{At
www.hpl.hp.com/econexperiment/Quantum~auction/instructions.htm}.
The subjects were told the decisions they were going to make, and
the potential consequences. They were also quizzed on the mechanics
of the game via the web site prior to coming to the lab for the
actual experiment. We implemented the experiments with the HP
Experimental Economics Software and followed standard procedures for
economics experiments.

After the subjects arrived in the laboratory and before starting the
experiment, we reviewed the instructions for the series of auctions
we would conduct during the experiment. We emphasized the subjects
will be bidding in groups of three and these groupings will be
randomized before each auction. This randomization was done in the
experimental software and did not involve any physical rearrangement
of the people in the room.

We described that subjects would be told their individual values for
winning the auction before the start of each auction. We further
told the subjects that a) their values are private meaning no other
players see them, b) the distribution of the values (uniform between
0 and 100) is known to everyone, and c) the values would be
randomized for every auction. To gain familiarity with the interface
and decisions, subjects played several training periods before the
actual experiment. In the training periods, subjects were not paid
for the outcome of the auction. The subjects were given about 90
seconds to make their decisions for each auction, although this rule
was relaxed somewhat in the training periods as well as the first
few periods of the game.

In the treatments with the quantum protocols, subjects were asked to
make \emph{three} decisions for each period: a bid (the amount they
had to pay if they won), and two numbers between 0-12. We first
explained to them that 0 is equivalent to 12. Thus, the two numbers
each behave like hours on a clock. Second, we said these two numbers
affect the probability that the highest, second highest and the
lowest bid wins the auction, as well as the probability for no
winner. Third, we described the decision support software tool that
calculates the outcome probabilities given their decisions, and
assumed values for their opponents' decisions. Using this tool, a
subject could enter a trial decision, i.e., values for the bid, $x$
and $y$, and guesses of the decisions of the other bidders. The
computer would then display, in table form, the probability of
winning for each bidder and the probability of there would be no
winner. \fig{screenshot} shows the interface people used to make
their decisions and evaluate consequences of possible choices.

For the experiments with the Stanford students, we did not describe
the underlying physics or present equations relating the outcome
probabilities to their choices. The subjects learn about the
relationships between their decisions and the actual probability of
winning by the use of the decision support tool. In the experiment
involving physicists we also described the quantum procedure
underlying the auction and how the standard and permuted search
treatments corresponded to different quantum search procedures.

The end of the experiment and upcoming switches to a different
treatment were announced two periods in advance, a standard protocol
in such experiments.

\begin{figure}
\begin{center}
\includegraphics[width=4in]{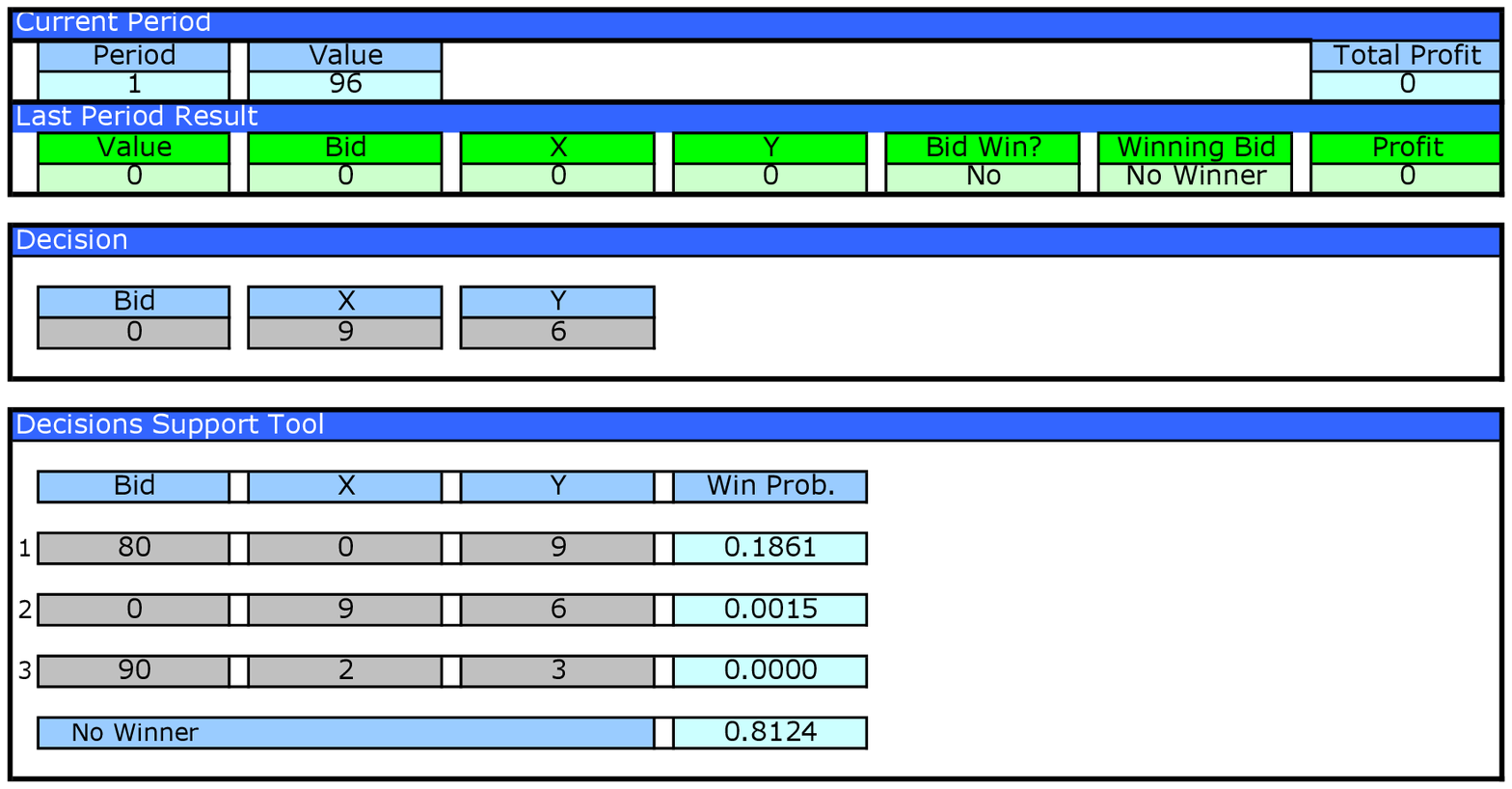} 
\end{center}
\caption{\figlabel{screenshot}User interface for making decisions.
Another sheet in the interface (not shown) listed the history of
outcomes during the game. The user entered choices for bid, $x$ and
$y$ in the ``Decision'' section. Prior to submitting their decision,
they could investigate consequences of different choices they and
others might make in the support tool at the bottom, which showed
outcome probabilities corresponding to those choices.}
\end{figure}

\section{Results}

Our experimental results allow quantifying and comparing the
performance of the three auctions. This section first describes
several common measures of auction performance and then uses these
measures to present the experimental outcomes.

\subsection{Auction Evaluation Criteria}

An key measure of auction outcome is the revenue, i.e., the amount
the seller receives from the auction. From the bidders' perspective,
the corresponding measure is their expected payoff, i.e., item value
minus amount bid if they win, and zero otherwise.

Allocative efficiency is a standard measure of the efficiency of an
economic mechanism. Instead of measuring how much money the system
generates (revenue), it measures whether the item is allocated to
the person with the highest value. Thus, it is defined as the ratio
of the achieved value to the maximum possible value. In the case of
auctions, allocative efficiency is simply the ratio of the value of
the winning bidder to the highest value amongst the bidders. Notice
that a high allocative efficiency does not necessarily imply high
revenue. For auctions with no winner, the allocative efficiency is
defined to be zero, corresponding to a situation in which there is
no value for the seller keeping the item.

Bidders must balance a low bid, giving higher profit if they win,
with a high bid, giving a higher chance of winning. Bidders with
high value can make a profit even with a relatively high bid. Thus a
measure giving insight into bidder strategies is the ratio of amount
bid to the item value for each bidder.

\subsection{Experimental Outcomes}

We evaluated the auction outcomes with several criteria, listed in
\tbl{results}. We report separately the averages over all auctions
and bidders in each treatment for each experiment (i.e., over all
periods and all groups of three in each period).

\begin{table}[t]
\begin{center}
\begin{tabular}{clccccc}
      &            & mean      & allocative & bid to value & mean   & fraction \\
expt. & treatment  & revenue   & efficiency & ratio        & payoff & no winner\\
\hline
1  & classical      & 58(2) & 97(1)\% & 75(2)\% & 4.3(0.4) & N/A\\
   & q. standard    &  7(2) & 34(4)\% & 18(2)\% & 6.1(0.7) & 51\%\\
   & q. permuted    & 14(2) & 33(4)\% & 34(3)\% & 3.0(0.5) & 59\%\\ \hline
2  & classical      & 56(2) & 97(1)\% & 89(16)\% & 4.7(0.5) & N/A\\
   & q. permuted    & 10(3) & 36(5)\% & 33(4)\% & 5.5(0.9) & 49\%\\ \hline
3  & classical      & 61(1) & 97(1)\% & 87(8)\% & 4.5(0.3) & N/A\\
   & q. standard    &  4(2) & 31(4)\% & 14(2)\% & 6.5(0.9) & 57\%\\
   & q. permuted    &  8(2) & 27(4)\% & 32(4)\% & 3.6(0.7) & 66\%\\ \hline
\end{tabular}
\end{center}
\caption{Auction performance measures. Averages are over the
auctions in the corresponding experiment and treatment. Numbers in
parentheses are the standard error in the estimate of the average
under the assumption of independence among the auctions. The
fraction of no winner applies only to the quantum auctions and
indicates the fraction of auctions that resulted in none of the
bidders winning the item.}\tbllabel{results}
\end{table}

As context for our experiments, the results of the classical
auctions are consistent with previous studies. For example, subjects
bid above 80\% of their values, as also seen in prior
experiments~\cite{cox88,chen98}. This similarity with prior
experiments suggests our experimental setup and subject pool are
sufficient to obtain behavior based on the economics forces in the
mechanisms.

A point of comparison for our observations with classical auctions
is with the predictions of game theory with risk-neutral bidders.
For the uniform distribution of values we use, theory predicts the
equilibrium bidding strategy with $\nBidders$ bidders has a bid to
value ratio of $(\nBidders-1)/\nBidders$, which is $66.7\%$ for all
our experiments since bidders competed in groups of three.
Furthermore, with uniform values between 0 and 100, the expected
largest value in a group of three is 75, with a corresponding bid
equal to 2/3 of this value, i.e., 50. Thus this theory predicts
average revenue for the classical auction is 50, a bit lower than we
observe experimentally. In fact, people are usually risk-averse,
which theory indicates leads to higher bids. Furthermore, game
theory predicts the allocative efficiency should be 100\% since the
highest bid always wins and the predicted equilibrium bidding
strategy has bids increasing monotonically (in fact, linearly) with
a bidder's value. However in practice, bidders vary somewhat from
this ideal bidding behavior. Thus occasionally a person with lower
value outbids the person with highest value. This situation is rare
in our experiments as seen by the high allocative efficiency ($\sim
97\%$) in all experiments. Similarly, theory predicts the average
payoff to the bidders: in this case 25 to the winner (75 average
value minus the bid of 50) and zero payoff to the two losing
bidders, for an overall average payoff per bidder of $25/3=8.3$.
This payoff is higher than the values we observe ($4.3$ to $4.7$),
again because people are bidding somewhat higher than theory
suggests.

Several conclusions on the quantum auctions can be drawn from the
experimental data. First, both types of quantum auctions, with the
standard search and permuted search, resulted in significantly lower
revenue and allocative efficiency than the classical auction. The
average revenue per auction dropped from around 60 in the classical
auction to less than 10 for the standard search treatment and around
10 to 15 for the permuted search. Specifically, pairwise Wilcoxon
tests comparing the classical auction with each of the two quantum
auctions indicate the differences in revenue, bid to value ratio,
mean payoff and allocative efficiency are statistically significant
with $p$-values less than $10^{-4}$. We can understand the lower
allocative efficiency as due largely to the significant fraction of
quantum auctions with no winner (since these cases give zero
efficiency). The many cases of no winner combined with the lower
bids leads to the lower observed revenue.

Second, comparing the two quantum auctions, the permuted search
resulted in significantly higher revenue and bid to value ratio
compared to the standard search. Pairwise Wilcoxon tests indicate
these differences are statistically significant with $p$-values less
than $10^{-4}$. The better performance of the permuted search
protocol is consistent with theory~\cite{harsha06}. In particular
the theory indicates the standard search gives bidders an incentive
to attempt to win with \emph{low} bids, whereas the permuted search
reduces this incentive by requiring collusion among bidders to
arrange for low bids to win (instead of having no winner).
\tbl{winning auctions} shows the permuted search has a much larger
fraction of winning auctions that are won by the highest bidder.
This difference between the two methods is significant ($p$-value
less than $10^{-4}$). Due to the lower bids in the standard search,
the mean payoff to bidders is higher in the standard search than the
permuted search (Wilcoxon test $p$-value $0.02$). However, the
fraction of auctions with no winners in the two treatments is not
significantly different ($p$-value of the proportional test $0.29$),
nor is the allocative efficiency ($p$-value of pairwise Wilcoxon
test $0.42$).

\begin{table}[t]
\begin{center}
\begin{tabular}{lccc}
treatment  & highest   & middle & lowest\\
\hline
q. standard    & 33\% & 21\% & 46\% \\
q. permuted    & 56\% & 11\% & 33\% \\
\end{tabular}
\end{center}
\caption{For auctions with a winner, fraction won by highest, second
highest and lowest bids for the standard and permuted search quantum
protocols.}\tbllabel{winning auctions}
\end{table}

Third, we can dispel the notion that players were entering random
$x$ and $y$ values because they were confused. The observation that
behavior changes systematically between the standard and permuted
search methods indicates users respond to the change in treatment.
Furthermore, the bid to value ratios were systematically below the
$0.5$ ratio that would arise from selecting bids uniformly at random
up to the bidders' values. Instead, the low ratios suggest people
were often attempting to win with a low bid by exploiting the
possibility that the highest bid is not always the winner. Finally,
the distribution of which bids won in auctions with a winner
(\tbl{winning auctions}) is significantly different ($p$-value less
than $10^{-4}$) from a uniform distribution that would arise if
auctions were won randomly.
These observations are strong evidence that subjects were developing
reasonable understanding of the implications of their decisions and
were responding to the strategic opportunities of the auction
protocol.

The large number of quantum auctions with no winner contributes
significantly to the lower performance we observed. Thus one key
challenge for the quantum auction is enabling coordination among
bidders to avoid choices leading to no winner (which give zero
profit to all bidders). To quantify the effect of this coordination
problem, \tbl{results winner only} shows the behavior for the subset
of quantum auctions that had a winner. For this subset of auctions,
revenue, allocative efficiency and payoff to the bidders are all
higher, as they must be since auctions without a winner contribute
zero to these measures. The bid to value ratio is also higher.
Nevertheless, these values still differ from the classical auction
($p$-value less than $10^{-4}$).
From this we conclude that coordination difficulties account for
some, but not all, of the difference in performance between the
classical and quantum auctions. In particular, even when bidders
manage to coordinate, their strategy still involves bidding fairly
low, either based on expectation that others also will bid low or in
the hope of winning in spite of not having the highest bid.

Comparing the two quantum auctions for cases with a winner, we find
significantly higher revenue and bid to value ratio in the permuted
vs.~the standard search ($p$-values less than $10^{-4}$). There is
no significant difference in allocative efficiency of the two
methods among winning auctions ($p$-value 0.15).

By our measure of bidder strategy, the bid to value ratio, we find a
significant difference ($p$-values less than $0.01$) in the ratios
involved in auctions with vs. without a winner when using the
standard search method. For the permuted search, ratios are higher
but the difference is not statistically significant. Thus, at least
for the standard search, users appear to adjust their bidding
strategy based on a sense of how much they expect to coordinate with
others to achieve a winning auction.

\begin{table}[t]
\begin{center}
\begin{tabular}{clcccc}
      &            & mean      & allocative & bid to value & mean   \\
expt. & treatment  & revenue   & efficiency & ratio        & payoff \\
\hline
1  & q. standard    & 14(3) & 69(4)\% & 21(4)\% & 12.6(1.1) \\
   & q. permuted    & 35(4) & 80(4)\% & 38(5)\% & 7.4(0.9) \\ \hline
2  & q. permuted    & 20(4) & 73(5)\% & 35(6)\% & 11.1(1.4) \\
\hline
3  & q. standard    &  8(2) & 72(5)\% & 20(4)\% & 15.3(1.5) \\
   & q. permuted    & 25(4) & 80(5)\% & 36(9)\% & 10.7(1.5) \\ \hline
\end{tabular}
\end{center}
\caption{Auction performance measures for the quantum auctions that
had a winner. Numbers in parentheses are the standard error in the
estimate of the average.}\tbllabel{results winner only}
\end{table}

\section{Conclusions}

In summary, we conducted a series of experiments to determine if
individuals can bid reasonably in a quantum auction protocol. We
conclude that subjects can react to the strategic consideration of
the auctions. In particular, the quantum protocol with permuted
search outperformed, with respect to efficiencies and revenue, the
one with standard search. This is consistent with prior game theory
analysis.

However, both treatments with quantum auctions resulted in
substantially lower revenue and efficiencies than classical first
price auction. This was largely driven by a much higher percentage
of ``no winner'' situations when subjects tried to manipulate the
winning probabilities through their choices of $x$ and $y$. Since
these cases of no winner can be avoided with suitable choices of $x$
and $y$, an interesting question is whether with further experience
participants would learn to avoid these no winner situations and
thereby increase their profits.

There are several directions for future work. First, the fact that
subjects responded to mechanism changes opens the possibility of
redesigning the protocol to improve efficiencies and revenue.
Second, this research has only focused on the quantum auction
protocol in isolation. It would be interesting to analyze and
conduct experiments with the same protocol in a larger economic
context (for example, where information about bids can be used in
future auctions) to see if the protocol has any benefits over
traditional auctions. The current experiments lack such context, but
indicate the quantum auction will only be economically beneficial if
its other properties (such as information privacy over repeated
interactions among the same group of bidders) outweigh their lower
revenue and efficiency.

We found that while quantum auctions were not as efficient as
classical auctions, subjects were responsive to design differences
in the quantum protocol. The experimental evidence suggests
participants had a fairly good understanding of the consequences of
their decisions and responded to structural changes in the quantum
protocol accordingly.

\section*{Acknowledgments}

We have benefited from discussions with Ray Beausoleil, David
Fattal, Saikat Guha, Pavithra Harsha and Cecilia Zenteno. This work
was supported by DARPA via the Army Research Office contract
\#W911NF0530002. This article does not necessarily reflect the
position or the policy of the Government funding agencies, and no
official endorsement of the views contained herein by the funding
agencies should be inferred.

\end{document}